# An Alternative Approach to Explaining Magnetospheric Processes

*E.A. Ponomarev, P.A. Sedykh*

*Institute of Solar-Terrestrial Physics SB RAS, Irkutsk, Russia*
pon@iszf.irk.ru , pvlsd@iszf.irk.ru

This paper gives a brief outline of the progression from the first substorm model developed in [Ponomarev, 1985; Sedykh, Ponomarev, 2002] based on C.F. Kennel's ideas, to the present views about the mechanism by which solar wind kinetic energy is converted to electromagnetic energy at the Bow Shock and by which this energy is transferred to the magnetosphere in the form of current; about the transformation of the energy of this current to gas kinetic energy of convecting plasma tubes, and, finally, the back transformation of gas kinetic energy to electromagnetic energy in secondary magnetospheric MHD generators.

The questions of the formation of the magnetospheric convection system, the nature of substorm break-up, and of the matching of currents in the magnetosphere-ionosphere system are discussed.

## 1. INTRODUCTION

We suggest a description of magnetospheric processes in the form of a physical model consisting of three blocks(see Fig.1):

1. Block of electric current generation in the Bow Shock where the Solar Wind (SW) energy converts to electric energy. The direction of the generated electric current depends on the sign of the IMF Bz-component. This current closes through the magnetospheric body in the form of the dawn-dusk current (if Bz<0). It is this current which sets convection in motion(by an Ampere force).

2. Block of gas pressure relief formation.
   The combined action of convection and pitch-angle diffusion [Kennel, 1969] leads to the formation of the gas pressure relief and, hence, to the production of a system of bulk and field-aligned currents in the magnetosphere, as well as to the formation of particle precipitation regions in the form of an oval corresponding to the location of the auroral zone. The nonstationary solution of this problem (with time-dependent boundary conditions) reproduces the substorm.

3. Block of magnetosphere-ionosphere coupling.
   The gas pressure relief determines the position of the MHD generators and the MHD compressor. It is shown that currents of MHD generators partially feed the ionospheric electric circuit and partially close on the MHD compressor. Such a structure of currents makes sit possible to reconcile the operation of magnetospheric sources of power and the ionospheric consumer.

Each block is realized in the form of a system of magnetic hydrodynamic equations taking into account mass and energy losses.

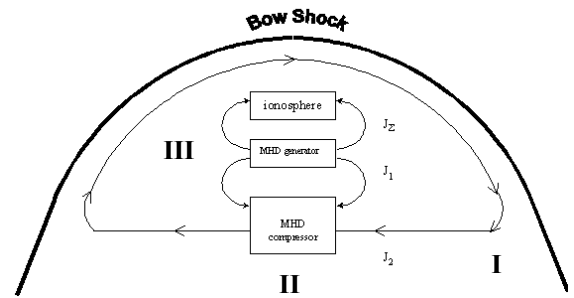

**Figure 1.** Layout of the functional blocks in the magnetosphere:
I - MHD generator that converts solar wind kinetic energy to electromagnetic energy;
II - MHD compressor that converts electric energy to gas pressure;
III - secondary MHD generators that convert compressed gas energy to electric current feeding electrojets in the ionosphere.

## 2. BLOCK OF ELECTRIC CURRENT GENERATED IN THE BOW SHOCK

The solar wind undergoes the greatest change of its parameters during the passage through the shock wave front. Its density in this case increases by a factor of four, and gas and magnetic pressures increase more than an order of magnitude.

The Bow Shock (BS), separating the Solar Wind (SW) region from the Transition Layer (TL) is, as shown in [Ponomarev et al., 2000,2003], a transformer that converts the solar wind kinetic energy to electric energy. As a result, under the shock wave there arises a current sheet separating the region of the Interplanetary Magnetic Field (IMF) of the solar wind from the magnetic field of the transition layer. If the vertical component of the IMF is in the solar ecliptic coordinate system, $B_z < 0$, then the current under the BS flows in the clockwise direction.

Fig. 2 shows a portion of the bow shock, the transition layer, and of the fore part of the magnetosphere.

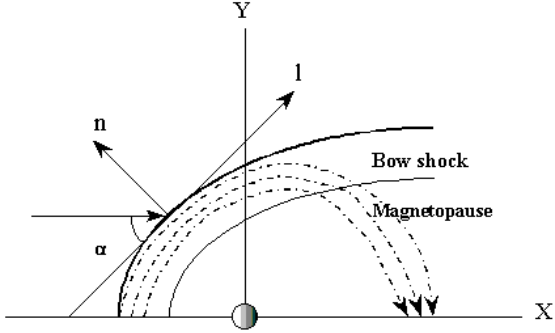

**Figure 2.** The position of the bow shock (BS), transition layer and magnetopause (MP). Schematically shown are the density lines of the electric current closing through the magnetosphere.

We now briefly discuss the conditions, which must be satisfied for the penetration of current into the magnetosphere. Let at the initial time the electric current be homogeneous and directed along the axis y. If at a certain time the current $j_o$ increases by $\delta j$ outside the volume under consideration, then a charge with surface density $\mu = \int \delta j\, dt$ starts to form on its boundary. The resulting electric field E will give rise to a displacement current $j_c = (\varepsilon/4\pi)\cdot\partial E/\partial t$, that creates an Ampere force which is balanced only by the inertial force, because the corresponding pressure gradient has not yet formed:

$$\rho_o\, \partial v/\partial t = [\mathbf{j_c} \times \mathbf{B}]/c \quad (1)$$

In magnetospheric conditions dielectric permittivity of plasma $\varepsilon = c^2/V_A^2$, then integrating (1) gives:

$$\mathbf{v} = c[\mathbf{E} \times \mathbf{B}]/B^2, \quad (2)$$

(in view of the fact that the squares of the perturbed quantities can be neglected as having the second order of smallness).

The physical meaning of what has been said above implies that the polarization electric field is produced inside the double layer with a thickness on the order of $\xi = 2\pi c_s/\omega_B = \pi c/\omega_{pp}$, where $c_s$ is the velocity of the fast magnetosonic wave, and $\omega_{pp}$ is the proton plasma frequency (such a thickness is characteristic for current sheets in collisionless laboratory and space plasmas). In the buildup process this field produces a displacement current that forms an Ampere force, which accelerates the plasma inside this layer. Whereas in the buildup process the electric field in the plasma coordinate system drops to zero and in the laboratory coordinate system, on the contrary, it increases to $[\mathbf{V} \times \mathbf{B}]/c$.

Because after the relaxation of the field inside the layer the boundary with surface charge has displaced into the plasma (in a time $t_1 \sim 2\pi/\omega_B$) to a distance on the order of $\xi$, all that has been described above is repeated. Obviously, this signifies the penetration of the electric field-associated momentum into the plasma with the velocity of sound. Here is how convection starts to form (see Fig. 3).

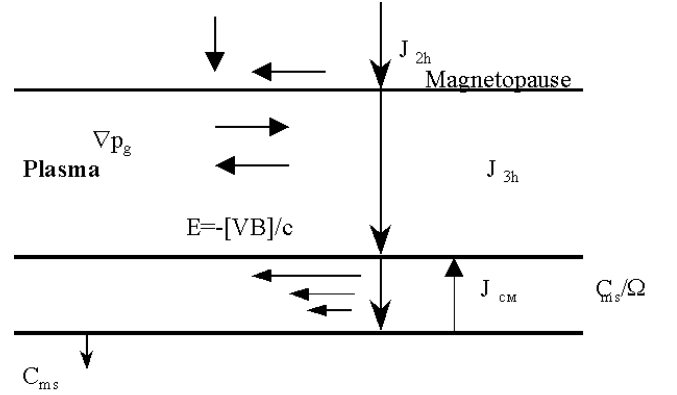

**Figure 3.** Schematic illustrating the penetration of the external current into plasma, and excitation of magnetospheric convection.

If there are no losses in plasma and div$\mathbf{V}=0$, then the story ends up with this. The plasma receives the momentum from the electric field, corresponding to the velocity of steady-state convection.

No stationary electric current is produced in this case. But if the volume is bounded by the walls or if it contains an inhomogeneous magnetic field, then primary convection undergoes restructuring, with the possible formation of a pressure gradient, the existence of which implies that an electric current arises in plasma. This does implies the penetration of the external current into plasma.

Next, we consider the formation of a pressure gradient caused by plasma compressibility. From the continuity equation we have:

$$\rho' = -\rho_o \int \text{div } \mathbf{V}\, dt \quad (3)$$

The integration goes over the entire time of convection formation $t_2$. Therefore (3) may be written as:

$$\rho' = \rho_o (V\nabla p_B/p_B)t_2 \sim \rho_o(V/L_B)t_2 \quad (4)$$

The perturbation of pressure $p'$ is determined from the equation of state:

$$p' = c_s^2\, \rho', \quad (5)$$

here $c_s$ is the velocity of the fast magnetosonic wave, since magnetic elasticity is of utmost importance in the magnetosphere.

In view of the fact that the velocity of fast magnetosound in the magnetosphere is virtually equal to the Alfvén velocity, we find:

$$\nabla p' \sim -\rho_o c_s^2 V t_2/L^2 \sim F, \qquad (6)$$

where F is the Ampere force. Since $V \sim F t_1/\rho_o$, from (6) it follows that:

$$t_1 t_2 = L^2/c_s^2 \qquad (7)$$

This relation relates the size of the system to the settling time of a stationary distribution of pressure, i.e. to the formation of a gas pressure gradient. This is known to determine the current in plasma.

In other words, $t_2$ is the characteristic time of penetration of the electric current into the plasma volume. Assuming for $t_1 = \omega_B^{-1} \sim 1$ s, $L \sim 10^{10}$ cm, and for $c_s \sim 3 \cdot 10^8$ cm/s, then for $t_2$ we find the time of ~1000 s. Such a duration of the transient process for the magnetosphere is unobjectionable. It turns out that our system has three characteristic times: $t_1 = 2\pi/\omega$ that characterizes the time of electric field buildup "at a point", $t_2$ is the time of penetration of the electric field into the volume, and $t_3 = L/c_s$ is the settling time of the electric field in the system of the size L. It should be noted that the gas pressure gradient is also produced when the magnetic field is homogeneous but there is a wall that confines the motion of plasma. Furthermore, the expression (7) retains its form, and the time $t_2$ acquires an illustrative character. It is simply the filling time of the volume between the wall and the point that is at a certain distance from the wall, with plasma moving with the velocity V toward the wall.

What has been said above suggests an important physical conclusion. The process of penetration of current into plasma is a two-stage one. Initially, the polarization field is produced, which penetrates into plasma "layer by layer". Or, more exactly, the momentum corresponding to this field penetrates into plasma. Here, if the system is inhomogeneous, the flow can redistribute pressure in such a manner that an electric current arises in plasma because of the appearance of gradients. Energetically, this current is necessary for maintaining convection in the inhomogeneous system. In fact, from the relation:

$$\mathbf{V}\nabla p = \mathbf{jE}, \qquad (8)$$

it follows that current is necessary for maintaining flow in an inhomogeneous medium. If (8) is integrated over the volume of the system, we see that in a stationary case the consumed power:

$$W = \int_U \mathbf{V}\nabla p \, dU = \int_U \mathrm{div}\mathbf{S} \, dU = \int_\Sigma \mathbf{S} \, d\Sigma = \int_\Sigma \psi \mathbf{j} \, d\Sigma, \qquad (9)$$

where U and $\Sigma$ are the volume and surface area of our system, and $\Psi$ is the surface potential. It is obvious that if the surface is closed and equipotential, i.e. $\psi$ can be taken outside the integral sign, then the system is energetically isolated from its surroundings - no energy can be "pumped in" (or "pumped out"), since $\psi \int_\Sigma \mathbf{j} d\Sigma = 0$. This was demonstrated by Heikkila [1997]. We use (9) to estimate the power brought into the magnetosphere by our current, with an appropriate potential difference. When $\Delta\psi \sim 600$ CGSE, $j_B \sim 5 \cdot 10^{-5}$ CGSE [Ponomarev et al.,2000] and $\Sigma \sim 10^{20}$ cm$^2$, for W we obtain the value on the order of $3 \cdot 10^{18}$ ergs/s, which corresponds to the value adopted for the perturbed state of the magnetosphere.

We have repeatedly stressed that equation (8) has a profound physical meaning. If $\mathbf{Ej}>0$, the electric forces do work on plasma. Plasma in this case moves toward increased pressure, that is, it is compressed. Otherwise the expanding gas produces an electric power. In the magnetosphere there are MHD compressors as well as MHD generators (see Fig. 1) [Ponomarev, 2000; Sedykh, Ponomarev, 2002]. If their combined output were identical, then the energy balance would be zero and the magnetosphere would not need any external energy sources. In fact, inside the magnetosphere is a constant energy consumer, the Earth's ionosphere. It is known that the energy flux density through the surface is proportional to the electric field component tangent to this surface.

It is obvious that in this way, by changing the degree of nonequipotentiality of the magnetopause, the magnetosphere can control the energy supplied externally. And the importance of this issue is thus. One can imagine that the intensity of magnetospheric processes is determined by the power alone, which is "offered" by the external source. Yet it may also be assumed that there exists also some regulator that permits the passage only a certain part of the "offered" power. Further, if this regular is linked to the consumer, then we have a very stable natural system.

Thus, the bow shock can be a sufficient source of power for supplying energy to substorm processes. The direction of current behind the BS front depends on the sign of the IMF $B_z$-component [Ponomarev et al., 2000,2003]. It is this current which sets convection in motion. Any change in external current through the magnetosphere causes a convection restructuring within a time on the order of the travel time of the magnetosonic wave from the magnetopause to the center of the system, because the restructuring wave comes from both flanks.

## 3. BLOCK OF GAS PRESSURE RELIEF FORMATION

It is known that the combined action of convection and strong pitch-angle diffusion of electrons and protons is responsible for the formation of gas pressure distribution in the magnetosphere[Kennel C.F., 1969; Ponomarev E.A., 1985], that is, steady volume currents. The divergence of these volume currents brings about a spatial distribution of field-aligned currents, i.e. magnetospheric sources of ionospheric current systems. We now consider this issue in slightly greater detail. It is known [Ponomarev, 1985] that the contents of the magnetic flux tube (MFT) to be referred to as the plasma tube (PT) throughout the text, transfers from one MFT to another in the convection process without surplus and deficiency in the case where the field lines of the magnetic flux tube are equipotential ones. This idealization is quite realistic everywhere apart from polar auroras.

Then, as the PT drifting toward the Earth in a dipole field, its volume decreases in proportion to $L^{-4}$, and the situation is the reverse for density, while pressure increases in proportion to $\sim L^{20/3}$. However, the process of adiabatic compression is attended by the processes of PT depletion due to pitch-angle diffusion into the loss cone. This process is described by the factor $\sim \exp(-\int dt/\tau) = \exp(-\int dr/V_r\tau) = \exp(-\int r d\upsilon/V_\upsilon \tau)$. The external current must perform the work on compressing these plasma tubes(by the Ampere force). Thus gas pressure has a maximum on each line of convection. In accordance with the equation for $p_g$ [Ponomarev, 1985], we have:

$$p_g = p_g^0 \left(\frac{L_\infty}{L}\right)^{\frac{20}{3}} \exp\left(-\frac{5}{3}\int \frac{dr}{V_r \tau}\right) \quad (10)$$

Here $p_g$ is gas pressure, L is the L-coordinate, $r = LR_e$ is the distance to the Earth ($R_e$ being the Earth's radius), $V_r$ and $V_\upsilon$ are the radial and azimuthal components of the convection velocity of the equatorial trace of the plasma tube, respectively, and $\tau$ is the characteristic time of PT depletion due to pitch-angle diffusion. The initial pressure at a certain boundary $L_\infty$ was considered time-independent in [Kennel, 1969]. For reasons unknown, Kennel did not extended his model to the unsteady-state case. This was done by one of us in [Ponomarev, 1985, 2000]. A typical gas pressure pattern that results through the combined action of convection and loses, is depicted in Fig. 4a. It has the form of an amphitheater with a clearly pronounced maximum near the midnight meridian, and with a sharp earthward "break". This "break" received the name "Inner Edge of the Plasma Sheet", IEPS.

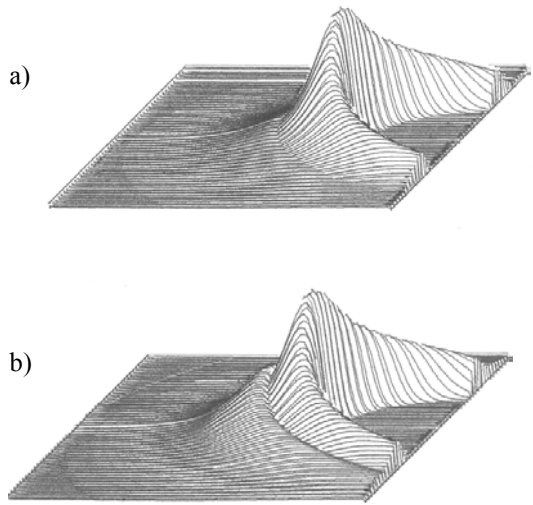

**Figure 4.** Gas pressure relief resulting from a combined action of plasma convection and losses as a consequence of particle precipitation into the ionosphere. Fig. 1b clearly shows a feature like a "gorge" which is produced during the influx of a plasma disturbance onto undisturbed pressure relief (as a result of the nonstationarity of boundary conditions).

The projection of the "amphitheater" onto the ionosphere corresponds to the form and position of the auroral oval(Fig. 5).

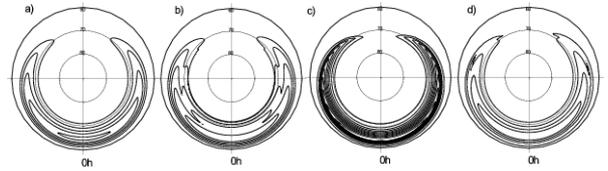

**Figure 5.** Equidensity contours of the precipitating electron flux for unsteady boundary conditions: a) t=0 s, b) t=1000 s, c) t=2800 s, d) t=4500 s.

This projection, like the real oval, executes a motion with a change of the convection electric field, and expands with an enhancement of the field. In this process the amplitude at a maximum increases as the IEPS approaches the Earth. Next we consider the case where the boundary conditions in (10) are time-dependent. Let the pressure on the boundary be increased by, say, a factor of two. This "impulse" will start to drift downstream with the convection velocity, with a region of double amplitude remaining everywhere in its wake. If the "impulse" is of short duration, then a region "multiplied by two" of a limited size will travel downstream. The effect of multiplication of two spatially narrow signals is always small apart from the time when their maximal coincide. An amplitude "flare" will occur then. Just this is the explanation for the "substorm breakup", a simple, logical corollary of the inhomogeneity of the system and motion of plasma [Ponomarev, 2000].

Fig. 4b illustrates the second phase of development of the pressure pattern in the process of a model substorm.

## 4. BLOCK OF MAGNETOSPHERE-IONOSPHERE COUPLING

Based on the spatial distribution of pressure as a function of coordinates and time, we can calculate the spatial distribution of volume currents:

$$\mathbf{j} = c[\mathbf{B}\mathbf{x}\nabla \mathbf{p_g}]/B^2 \quad (11)$$

The divergence (11) under steady-state conditions gives an expression for field-aligned current densities(Fig. 6):

$$j_\parallel = cB_I \int_0^l \{[\nabla \mathbf{p_g} \mathbf{x} \nabla \mathbf{p_B}] \cdot \mathbf{B}/p_B B^3\} dl \quad (12)$$

We perform the integration along a magnetic field line of the Earth's dipole field from the equator (0) to the ionosphere (l).

Noteworthy is the following property of the expression under the integral sign. It depends on the angle of intersection of magnetic and gas pressure contours. Within the dipole approximation $p_B$ = const are merely circles. On the contrary, $p_g$ = const have a complex configuration. The sign of current $j_{\parallel}$ depends, ultimately, on the sine sign of the angle between the normals to pressure contours. This factor eases qualitatively analysis of the current situation.

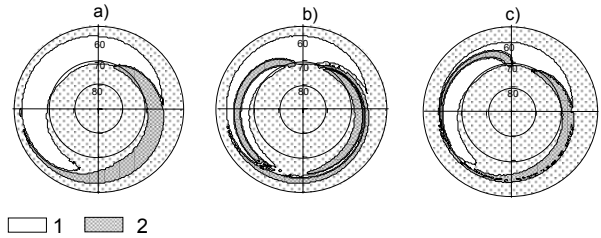

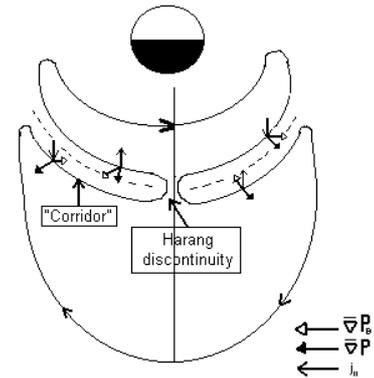

**Figure 7.** Schematic of the section of the gas pressure relief. The section of the "gorge" is represented by "corridors", on the walls of which field-aligned currents are generated.

**Figure 6.** Field-aligned currents generated in the magnetosphere: 1 - zone of inflow currents, 2 - zone of outflow currents; a) t=0 s; b) t=1000 s; c) t=2800 s.

The complexity of the magnetosphere-ionosphere coupling (MIC) problem implies that currents in the ionosphere are governed by the electric field (with conductivity specified as a parameter), and in the magnetosphere they are determined by gas pressure gradient. There does exist a connection between the pressure distribution and convection, albeit relatively complicated. Our intention is to understand (by analyzing a maximum possible simple model that at the same time retains the most important traits of reality) how consistently current is established in the overall ionosphere-magnetosphere chain, how the magnetospheric generator of ionospheric currents operates, and what sources of power (including those of no electromagnetic origin) this generator uses to be at work. A partial answer to the last question has been given to date. We have demonstrated [Ponomarev, 1985] that magnetospheric regions that operate like an MHD compressor where plasma is compressed under the action of Ampere's force $[\mathbf{j}\times\mathbf{B}]/c$, satisfy the condition $\mathbf{V}\cdot\nabla p_g>0$, and regions where gas dynamic forces acts on electromagnetic forces, i.e. regions of MHD generators, satisfy the condition $\mathbf{V}\cdot\nabla p_g<0$. Conversion of energy from one kind to another may be written by a straightforward formula:

$$\mathbf{V}\cdot\nabla p_g = \mathbf{j}\cdot\mathbf{E}$$

It seems appropriate to employ in the analysis the region of the "cleft"(or the "gorge") which is produced when a plasma disturbance flows against the undisturbed pressure pattern (as a result of the unsteady-state character of boundary conditions as mentioned above). This detail of the pattern is clearly seen in Fig. 4b. Fig. 7 shows a schematic representation of a section of this pattern.

The section of the cleft is represented by "corridors". One can see that the walls of "corridors" serve as the sources of two bands of field-aligned currents which direction is opposite on different walls. On the whole, a current configuration forms, which corresponds to the Iijima-Potemra scheme [Iijima and Potemra, 1976]. Importantly, the stream convection lines run virtually along the axis of the "corridor", the "corridor" itself is extended with respect to the $p_B$=const contours at a small angle, and hence the magnetic field inside it is nearly homogeneous. For that reason, the precipitation parameter $\tau$ can be considered a constant quantity.

In the model of our interest, we replace the "corridor" itself by a rectangular channel with perfectly conducting walls overlaid by a conducting "cover", the ionosphere. We compensate for the difference in spatial scales, which is caused by the convergence of field lines, by a correction of parameters. The channel with a homogeneous magnetic field includes a steady flow of ideal plasma with a corresponding pressure gradient. All this is portrayed in detail in Fig. 8.

Let us consider the phenomena occurring in the plasma "corridor" on the basis of a simple model. As is evident from Fig. 7, the orientation of the "corridor" is such that plasma flows nearly along its axis. The corridor is extended in a longitudinal direction; therefore, the magnetic field changes little within it. All these factors allow us to replace the "corridor" by a channel (extended along the axis Y) of width 2D, length L, and height H. The axis Y will be oriented across the channel, and the axis Z along its height, as shown in Fig. 8. The channel is filled with ideal plasma with pressure $p^0$ at the inlet and $p^1$ at the outlet. The magnetic field $\mathbf{B} = \{0,0,B_z\}$ will be considered homogeneous. Plasma with the velocity $\mathbf{V} = V_x(x)$ flows along the axis X in a positive direction. The walls of the channel possess infinite conductivity. The ionosphere is modeled by the upper cover of thickness h with Pedersen conductivity $\sigma$.

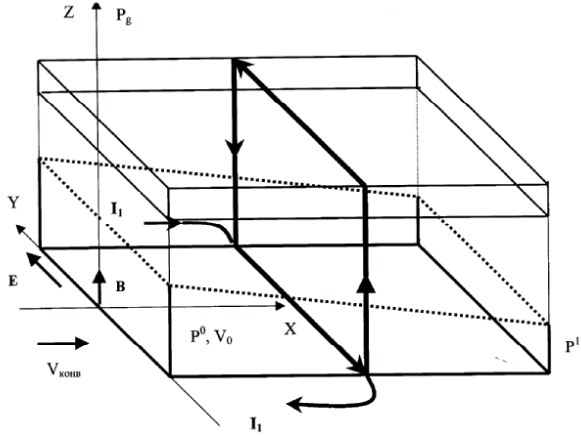

**Figure 8.** Schematic representation of the channel of the magnetospheric MHD generator with perfectly conducting walls overlaid by a conducting "cover", the ionosphere. The dashed line shows the gas pressure pattern $P_g$, and the thick lines show the direction of currents (for designations see the text).

As a consequence of the existence of a pressure gradient along the channel, the following current flows across it:

$$j_y = \frac{c}{B}\left(\frac{\partial p}{\partial x}\right)$$

To this volume density of current there corresponds the surface density and a total current:

$$I_G(x) = \int j_y dz = H j_y \ , \quad J_G = \int I_G dx$$

Accordingly, a total current of ionospheric load is:

$$J_\sigma = \iint \sigma E_I dx' dz' = \sigma h \int E dx = \frac{\sigma \cdot hB}{c}\int V dx \quad (13)$$

The primes on the differentials signify that the integration is performed over the space of the ionosphere. Furthermore, because of the equipotentiality of magnetic field lines, the electric field in the ionosphere $E_I$ is related to the electric field in the magnetosphere by the relation: $E_I dx' = E dx$. In these formulas, c is the velocity of light.

In addition to the current that closes through the ionosphere, a part of the MHD generator's current can close through the magnetosphere, as is the case with the corridor's current in Fig. 7. We designate this current by index 1. Then:

$$J_1 = \iint j_{y1} dx dz = \int I_1 dx$$

From the condition of continuity of currents we find:

$$\frac{dp}{dx} = -\frac{\sigma * B^2 V}{c^2} + \frac{I_1 B}{cH} \quad (14)$$

where $\sigma^* = \sigma(h/H)$.

The balance equation of gas kinetic energy in a steady-state one-dimensional case has the form:

$$V\frac{dp}{dx} + \gamma \cdot p \frac{dV}{dx} = -\gamma \frac{p}{\tau} \quad (15)$$

Whence:

$$p = p^0 \left(\frac{V_0}{V}\right)^\gamma \exp\left(-\gamma \int \frac{dx}{V\tau}\right) \quad (16)$$

We now designate the initial level of gas pressure that is necessary and sufficient for supplying the ionosphere with electric current, by $p^{01}$ so that $p^0 = p^{01} + p^{02}$, where $p^{02}$ is the initial level of gas pressure that produces a current $J_1$. Then:

$$\gamma p^{01}\left(\frac{V_0}{V}\right)^{\gamma+1} \exp\left(-\gamma \int \frac{dx}{V\tau}\right)\left[\frac{dV}{dx} + \frac{1}{\tau}\right] = \sigma *\left(\frac{B}{c}\right)^2 VV_0 \quad (17)$$

$$-\gamma\left[p^0 - p^{01}\left(\frac{V_0}{V}\right)^{\gamma+1}\right]\exp\left(-\frac{\gamma}{\tau}\int\frac{dx}{V}\right)\left[\frac{dV}{dx}+\frac{1}{\tau}\right] = I_1\left(\frac{B}{cH}\right)V_0 \quad (18)$$

The solution of this system of equations that satisfies the conditions of our problem, is:

$$V = V_0 - \frac{\gamma x}{(\gamma+2)\tau} \quad (19)$$

From (17) we obtain the condition:

$$p^{01} = \frac{(\gamma+2)}{2\gamma}\left(\frac{B}{c}\right)^2 \sigma * \tau \cdot V_0^2 \quad (20)$$

And from (18) we get:

$$I_1 = -\frac{2\gamma}{(\gamma+2)}\left[p^0 - \frac{(\gamma+2)}{2\gamma}\left(\frac{B}{c}\right)^2 \sigma * \tau V_0^2\right]\frac{VcH}{B\tau V_0^2} \quad (21)$$

It is evident from (21) that the current $I_1$ is "organized" by the "principle of balance": all the necessary expenses of the ionosphere in current (power) are covered first, and what remains leaves for the geomagnetic tail region. As is evident from the figures, the current $I_1$ there becomes part of the dawn-dusk current. Only a part because there exists also the dawn-dusk current $J_B$ of a different origin. It is an external current with respect to the magnetosphere itself. As was shown by Ponomarev et al. [2000], it is produce at the Bow Shock (BS) front through a partial deceleration of solar wind plasma by Ampere's force with the involvement of this current. If the $B_z$-component of the IMF is less than zero, the direction of this current is such that, by closing through the magnetospheric body, it produces there Ampere's force capable of acting to pushing magnetospheric plasma earthward, toward an increase of magnetic and gas pressure (see Block of electric current generation in the BS). Thus the MHD compressor lies in this region (located mostly at 5<L<10 on the nightside, i.e. before the gas pressure maximum). It is the gas compressed by the generator that is supplied to the MHD channel, the operation of which we are discussing here. Unlike the channel's region, the region of the MHD compressor lies in the area where the plasma is driven by magnetospheric convection to travel nearly radially to the Earth. From the balance of the gas pressure force and Ampere's force we have:

$$J_1 + J_B = cH \int B^{-1} \left(\frac{dp}{dL}\right) dL, \qquad (22)$$

where $B = B_0/L^3$.
Whence:

$$p^0 = q\left(\frac{B_c}{cH}\right)[J_1 + J_B], \quad q = \frac{(4\gamma - 1)}{4\gamma}\left(\frac{L_c}{L_T}\right)^{4\gamma} L_c^2, \qquad (23)$$

$L_c$ and $L_T$ are the coordinates of the end and beginning of the area of plasma compression, and $B_c$ is the magnetic field strength at the compressor output. Further it will be assumed that $B_c=B$, that is, the MHD compressor output territorially coincides with the MHD generator input.

Since plasma requires some time to travel the distance from the compressor input to output:

$$\Delta T = \int_{L_T}^{L_c} \frac{R_e dL}{V_R},$$

then pressure at the MHD generator input will correspond to the earlier value of the compressor current.

By integrating (21) over the entire length of the channel and assuming that the plasma velocity at the output is much smaller than that at the input of the MHD generator, we find:

$$J_1 = \left(\frac{cH}{B}\right)\left[p^0 - \frac{(\gamma+2)}{2\gamma}\left(\frac{B}{c}\right)^2 \sigma^* \tau V_0^2\right] \qquad (24)$$

Upon substituting (23) into (24), in view of what has been said about the delay, we obtain an important relation:

$$J_1(t) - qJ_1(t - \Delta T) = qJ_B(t - \Delta T) - J_\sigma(t) \qquad (25)$$

In a steady state where there is no explicit time-dependence and q = 1:

$$J_B = J_\sigma \qquad (26)$$

This means that actually dissipative processes can take place in the magnetosphere only at the expense of an external source of current (and energy).
The whole of the complicated magnetospheric "design" only redistributes currents and energy fluxes in space and time.
Overall, though, this is an obvious inference as it is expectable. The integrity of (26) in this case implies that this is not merely a declaration now. We can point out the limits of applicability of (26) as well as the particular processes behind the notions "steady state" and "unsteady state".

We now turn our attention to the "cross-tail currents". Let $J_1+J_B$ be designated by $I_s$. Then from (25) it follows that:

$$J_s(t) = qJ_s(t - \Delta T) + [J_B(t) - J_\sigma(t)] \qquad (27)$$

Obviously, the control of the tail current $I_s$ proceeds both at the expense of a variation of $J_B$ and at the expense of the variation of the current of ionospheric load $J_\sigma$. In a quasi-steady situation where $J^{-1}dJ/dt \ll 1$, q=1 we have:

$$\frac{dJ_s}{dt} \sim \frac{[J_B - J_\sigma]}{\Delta T} \qquad (28)$$

Obviously, when $J_B>J_\sigma$, the cross-tail current increases, and the magnetospheric magnetic field is observed to extend into the tail. Otherwise when the ionospheric load current exceeds the external current, $dJ_s/dt <0$ and the tail current decreases, a "dipolarization" of the magnetic field occurs. The physical reason behind this is the increase in ionospheric consumption of current because of the increase in of conductivity caused by an enhancement of auroral particle precipitation.

Thus between the consumer of current and energy, on the one hand, and their "general supplier", the external current, there exists a flexible connection via a "depot" represented by current $J_1$.

Two serious arguments can be adduced in favor of our developed magnetosphere-ionosphere coupling model. Indeed, the prototypes of the channels considered in this study, the plasma "corridors", lie in the immediate vicinity of the pressure maximum, i.e. of the particle precipitation maximum. It is precisely where electrojets are located. The other argument implies that the configuration of field-aligned currents in this case reproduces the picture of inflow and outflow currents (see Fig. 6), which was experimentally established in [Iijima, Potemra, 1976].

## 5. CONCLUSION

The approach under development allows an understanding of:
- How the energy is derived from the SW, which is needed for magnetospheric processes, and how it is transferred to the magnetosphere.

- How the break-up is produced. The shape and spatial displacement of the auroral oval depending on time. The location, structure and origin of auroral electrojets.
- How the magnetospheric currents can be associated with ionospheric currents dependent on the electric field and conductivity.

*Acknowledgement.* This work was done under RFBR project №. 02-05-64066.

REFERENCES

Heikkila W.J. Interpretation of recent AMPTE data at the magnetopause, *J. of Geophys. Res.*, 102, 2115-2124, 1997.

Iijima T., Potemra T.A. The Amplitude Distribution of Field-Aligned Currents at Northern High Latitudes Observed by TRIAD. // *J. of Geophys. Res.*, v.81, p. 2165-2174, 1976.

Kennel C.F. Consequences of magnetospheric plasma. // *Rev. Geophys.*, 7, p. 379-419, 1969.

Ponomarev E.A. The Mechanisms of Magnetospheric Substorms. *Moscow: Nauka*, 1985.

Ponomarev E.A. On one plausible simple explanation for substorm break-up, *Proc. 5th International Conference on Substorms, ESA SP-443*, 549, 2000.

Ponomarev E.A., Urbanovich V.D., Nemtsova E.I. On the excitation mechanism of magnetospheric convection by the Solar Wind, *Proc. 5th International Conference on Substorms, ESA SP-443*, 553, 2000.

Ponomarev E.A., Sedykh P.A., Mager O.V., Urbanovich V.D. Conditions of excitation of magnetospheric convection by the electric current generated in the bow shock. *LANL e-print archive.* http://arxiv.org/abs/physics/0306041 , 2003.

Sedykh P.A., Ponomarev E.A. Magnetosphere-ionosphere coupling in the region of auroral electrojets. *Geomagnetism and Aeronomy*, V.42, №5 P. 582-587, 2002.